# Eye Gaze Controlled Interfaces for Head Mounted and Multi-Functional Displays in Military Aviation Environment

LRD Murthy[1], Abhishek Mukhopadhyay[1,2], Varshit Yellheti[2], Somnath Arjun[1], Peter Thomas[3], M Dilli Babu[4], Kamal Preet Singh Saluja[1], JeevithaShree DV[1] and Pradipta Biswas[1]

| [1]Indian Institute of Science Bangalore, India | [2]Indian Institute of Information Technology Kalyani, India | [3]University of Hertfordshire Hertfordshire, UK | [4]Aircraft and System Testing Establishment Bangalore, India |
|---|---|---|---|

*Abstract*—Eye gaze controlled interfaces allow us to directly manipulate a graphical user interface just by looking at it. This technology has great potential in military aviation, in particular, operating different displays in situations where pilots' hands are occupied with flying the aircraft. This paper reports studies on analyzing accuracy of eye gaze controlled interface inside aircraft undertaking representative flying missions. We reported that pilots can undertake representative pointing and selection tasks at less than 2 secs on average. Further, we evaluated the accuracy of eye gaze tracking glass under various G-conditions and analyzed its failure modes. We observed that the accuracy of an eye tracker is less than 5º of visual angle up to +3G, although it is less accurate at -1G and +5G. We observed that eye tracker may fail to track under higher external illumination. We also infer that an eye tracker to be used in military aviation need to have larger vertical field of view than the present available systems. We used this analysis to develop eye gaze trackers for Multi-Functional displays and Head Mounted Display System. We obtained significant reduction in pointing and selection times using our proposed HMDS system compared to traditional TDS.

## TABLE OF CONTENTS

## 1. INTRODUCTION

Eye tracking is the process of measuring either the point of gaze where one is looking or the motion of an eye relative to the head. This paper investigated use of eye gaze trackers in military aviation environment as a direct controller of various types of user interfaces like Head Down and Head Mounted Display systems.

Presently, eye gaze tracking devices are easily available and have been used to directly control user interfaces. Eye gaze controlled displays are mainly explored for assistive technology to make computers accessible to people with severe physical impairment [5]. It also found application to facilitate interaction for touchscreen or mouse. In recent time, eye gaze controlled interfaces are explored for automotive user interfaces [4] and de Reus [9] proposed to use eye gaze trackers as a direct controller of Head Mounted Display System (HMDS). Use of eye gaze tracking to analyze pilots' interaction with cockpit displays dated back to 1950s [10, 19]. Eye tracking has already been used for flight safety in the following ways:

- Comparing scan paths and fixation durations to evaluate the progress of pilot trainees,
- Estimating pilots' skills,
- Analyzing crew's joint attention and shared situational awareness,
- Displaying a notification at the point of pilot's gaze to ensure its visual processing, performing an automatic maneuver and so on.

Eye gaze controlled interface has great potential for military aviation as pilots find it difficult to use existing target designation system in high G situations and direct voice input systems are not well explored for non-native English speakers and for languages other than English. Additionally, eye gaze trackers can also be used to automatically estimate pilots' cognitive load [12, 3]. However, eye gaze controlled interfaces need to be evaluated in actual flight conditions as earlier studies [6, 7, 9] only used them in simulators. Adelstein [2] and colleagues reported *"significant* *degradations in both error rate and response time in a reading task at 0.5 and 0.7 g for 10-pt, and at 0.7 g for 14-pt font displays*".

In this paper, we first study the effectiveness of using eye gaze trackers for undertaking representative pointing and selection tasks in a transport aircraft during different phases of flight on a display in Head Down configuration. This configuration is similar to the set up of Multi-functional displays (MFDs) in actual flight cockpits. Next, we study gaze tracking accuracy and failure modes of eye gaze tracker under constant G manoeuvres in a BAES Hawk Trainer aircraft.

Based on our results and analysis, we present two eye trackers developed for a simulated HMDS using wearable eye tracker and for MFDs using a web camera. For HMDS, we develop a multimodal head and eye gaze tracking system and integrated it with a flight simulator. Our studies showed the system can statistically significantly reduce target locking duration compared to traditional TDS.

## 2. STUDY ON TRANSPORT AIRCRAFT

This study undertook an ISO 9241 pointing tasks inside a transport aircraft. In particular, we compared two different versions of eye gaze controlled interface – one version only moves pointer in a graphical user interface using eye gaze (non-adaptive), the other version not only moves the pointer but also activates target nearest to the eye gaze location (adaptive). Details on the pointer movement algorithms is described in a separate paper [12].

We collected data from 3 IAF (Indian Air Force) pilots with ranks ranging from squadron leader to wing commander. We collected data using a Microsoft Surface Pro tablet running Windows 10 operating system and a Tobii PCEye Mini eye gaze tracker [20]. The eye tracker has an accuracy of 0.4° of visual angle at desktop computing environment. We used an Avro HS748 transport aircraft for data collection purpose. We used a X16-1D USB Accelerometer from Gulf Coast Data Concepts for recording vibration in unit of $g$ (g =9.81 m /s$^2$). We set up the tablet and eye gaze tracker at the front seat outside the cockpit as shown in Figure 1 below.

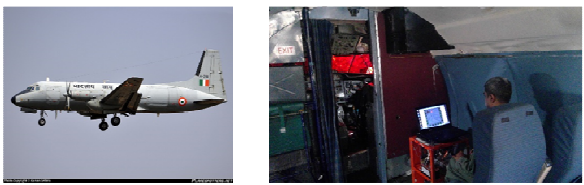

**Figure 1. Aircraft used in the study and placement of set up inside the aircraft**

We used the ISO 9241 pointing task with the following target sizes and distances.

- Target sizes ($W$, in cm): 1.9, 1.7, 1.5, 1.3, 1.1, 0.9
- Distance of target from center of screen ($D$, in cm): 5, 8

We designed a repeated measure study with following independent variables

- *Place of Study*
  - On Ground
  - In Air
- *Type of System*
  - Non Adaptive
  - Adaptive, with nearest neighborhood algorithm that activates target nearest to gaze location.

We also used an accelerometer in front of the tablet computer to record vibration while flying. In total, we analyzed 956 pointing tasks with at least 150 tasks recorded for each condition. We calculated average movement time for all combinations of width and distances to target for all different conditions. Figure 2 plots the movement times with respect of indices of difficulties ($ID = \log_2\left(\frac{D}{W}+1\right)$) for all four conditions. We found correlation coefficient r=0.64 and r=0.63 between movement time and ID for the non-adapted versions on ground and air respectively. However, with the nearest neighborhood algorithm, the correlation coefficients were less than 0.3.

We undertook a *Place of Study (2) × Type of System (2)* repeated measure ANOVA on the movement times. We found

- A significant main effect of *Place of Study* F(1,10) = 14.38, p<0.05, η$^2$ = 0.59
- A significant main effect of *Type of System* F(1,10) = 34.80, p<0.05, η$^2$ = 0.78
- An interaction effect of *Place of Study* and *Type of System* F(1,10) = 7.78, p<0.05, η$^2$ = 0.44

A set of pairwise comparisons found that there are significant differences at p<0.05 in movement times between data collected at ground and on air and between adapted and non-adapted conditions on data collected on air.

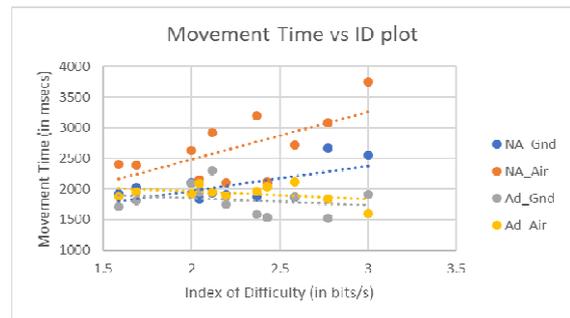

**Figure 2. Movement time vs ID plot**

In terms of qualitative feedback, all pilots preferred the adaptive version over the non-adaptive one. In particular, they noted that the non-adaptive version turns difficult to use during take-off and landing phases compared to cruising phase. In terms of application, they noted that the system will be useful for operating the MFD and operating the HMDS for investigating and engaging beyond visual range targets.

We further analyzed the cursor movement trajectories for both adaptive and non-adaptive conditions. Cursor movement efficiency can be analyzed in detail using the metrics defined by MacKenzie [14] with regards to the task axis (the line between starting source and intended target). These metrics look at the variability of movements as well as the number of events relating to cursor movement along the task axis towards the intended target. These are illustrated in Figure. 3.

Movement variability (MV) is defined as the standard deviation in the variation in orthogonal movement, relative to the average deviation:

$$MV = \sqrt{\frac{\sum_{i=1}^{n}(y_i' - \bar{y}')^2}{n-1}}.$$

Movement error (ME) concerns the overall mean magnitude of deviation in cursor movement from the task axis:

$$ME = \frac{\sum_{i=1}^{n}|y_i'|}{n}.$$

The movement offset (MO) is the average magnitude of deviation, $\bar{y}'$. Both the orthogonal and movement direction change metrics (ODC and MDC, respectively) indicate the number of times the participant changes the direction of cursor movement orthogonal to, and parallel to, the task axis, respectively, as the cursor is moved towards the intended target. The task axis crossing (TAC) concerns the number of times the cursor is moved across the task axis as the cursor is moved towards the target. Lastly, target re-entries (RE) counts the number of times the cursor moves back into the target area after first leaving.

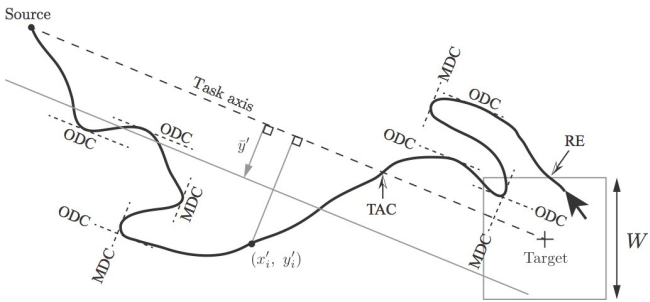

**Figure 3. Illustration of cursor efficiency metrics**

Initially we compared the average values of these parameters in adaptive and non-adaptive conditions (Figure

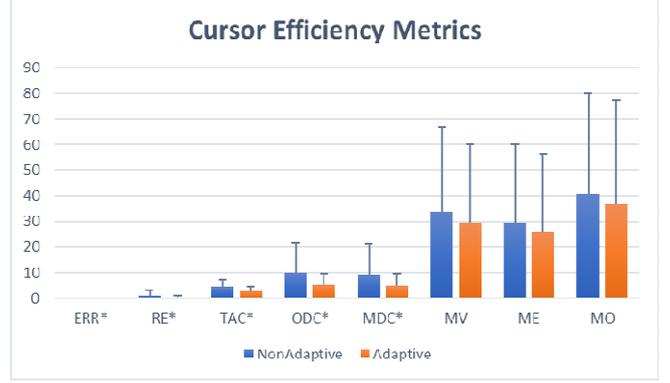

**Figure 4. Comparing cursor efficiency metrics between adaptive and non-adaptive conditions**

4). We used a * marks on the graph for parameters, which were significantly different at p<0.05 in an unequal variance t-test.

Furthermore, we analyzed the data with two more visualization techniques for obtaining better insight about the information, which is not feasible with the above plotted graphs. We examined with Radial Stacked Bar Chart and developed a novel approach called Scattered Radial Bar plots. Using Radial Stacked Bar Chart [1,18], we can view and compare all dependent variables for adaptive and non-adaptive data in one visual frame against all IDs together instead of plotting each dependent variable separately against each ID. From Figure 5 it may be noted that the extent of the gray and brown part that the avg_MO and avg_ME for target width 100 and distance 200 is larger than other IDs. We also observed that avg_ME for target width 100 and distance 300 is largest in adaptive data but in non-adaptive data target width 90 and distance 300 has the largest avg_ME. In non-adaptive part of the figure, effectiveness of all the dependent variable cumulatively for target width 90 and distance 300 is the largest amongst all values of IDs and for the adaptive section of the visualization the overall effectiveness of all the dependent variable together for target width 50 and distance 350 and target width 100 and distance 300 are the largest despite the time taken to complete the task being different for both of them. From the graphs, we can note that ME and MV have different values for different IDS while RE and Error were not much affected by different values of IDs. In Scattered Radial Bar plots [8,18], we examined dependent variables for each IDs individually, for example in Figure 6 we can comprehend that avg_ERR for target width 90 and distance 200 is more compared to other IDs, which is evident from the green part of the bar in the plots. We also noticed that avg_MO, avg_ME, avg_MV which is represented by gray, brown and pink color respectively for all the IDs are almost similar irrespective of the time.

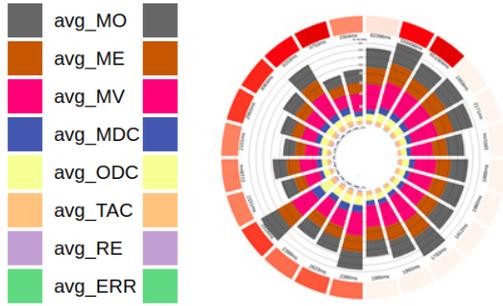

**Figure 5. Radial Stacked Bar Chart**

We separately analyzed the vibration profile (Figure 7) in terms of the acceleration values recorded for roll, pitch and yaw. The roll and yaw vibration had a maximum value of 1.2G while the acceleration measured for pitch reached 1.5G.

This study shows that the nearest neighborhood algorithm made selection of smaller targets easier as indicated by the low value of correlation between movement time and ID. This ease of selection of small targets turn more useful on air under vibrating condition than on ground as indicated by the ANOVA study and pairwise comparisons. It may also be noted that using the nearest neighborhood algorithm, participants can select target using gaze controlled interface in less than 2 secs on average on both ground and air. Analysis on the cursor efficiency metrics show that the nearest neighborhood algorithm reduces the homing time to target as indicated by the significantly lower values of RE, TDC, ODC and MDC but at a cost of higher error rate, although it is less than 5%.

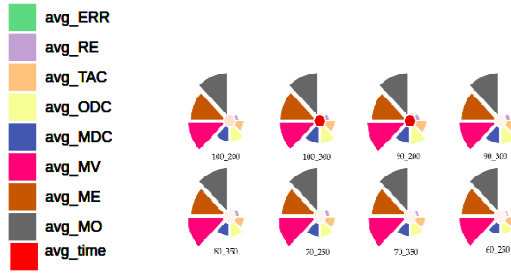

**Figure 6. Scattered Radial Bar**

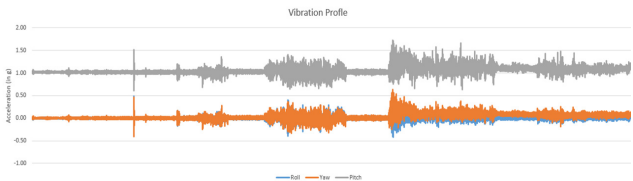

**Figure 7. Sample vibration profile of the aircraft**

This study shows that the nearest neighborhood algorithm made selection of smaller targets easier as indicated by the low value of correlation between movement time and ID. This ease of selection of small targets turn more useful on air under vibrating condition than on ground as indicated by the ANOVA study and pairwise comparisons. It may also be noted that using the nearest neighborhood algorithm, participants can select target using gaze controlled interface in less than 2 secs on average on both ground and air. Analysis on the cursor efficiency metrics show that the nearest neighborhood algorithm reduces the homing time to target as indicated by the significantly lower values of RE, TDC, ODC and MDC but at a cost of higher error rate, although it is less than 5%.

However, in this study, we used two different devices for measuring movement time and acceleration and hence cannot synchronize it in milliseconds level. We could not make separate analysis for different flying phases and being in a transport aircraft, we could measure performance of the gaze controlled system up to 1.5G only. In our future studies, we are planning to collect data on a fighter aircraft attaining higher G values and syncing the users' performance with vibration profiles.

### 3. STUDY ON FIGHTER AIRCRAFT

In this study, we collected data from one pilot (age 35 years, flying experience of 1920 hours in multirole combat aircraft) in constant G manoeuvre level turns in a BAES Hawk Trainer aircraft. The BAE Systems Hawk is a British single-engine, jet-powered, twin seater trainer aircraft in tandem seating configuration. We collected data at +5G, +3G, -1G and compared with data collected at +1G. A demonstration video can be found at https://cambum.net/ConstantG.mp4 .

We used the Tobii Eye Tracking glasses [21] for data collection, it has one scene camera to record outside view and four cameras, two for each eye record eye gaze at 100Hz. A proprietary software (Tobii Pro Studio) maps eye gaze on the video recorded in the scene camera and indicates the point of gaze fixation by drawing a red circle on the scene video. The recorded point of fixation is referred as gaze point in subsequent analysis.

We have analyzed the accuracy between gaze point and target point from the videos recorded in different G values by calculating the distance between target point and gaze point distance. We have used image processing methods (Figure 8) for this measurement. Initially, we did color transformation and removed noise in the image and applied adaptive threshold to find region of interests as written in **Algorithm 1**. This image is given as input to **Algorithm 2**. We have applied Hough transform to find both target and gaze circles in the preprocessed image. It returns Euclidean and Manhattan distance in units of pixels (Figure 9). To convert the pixel distance into centimeter, we measured radius of target circle in photo, that is 2.2 cm. We measured the area of the contour around the target circle, thus measured radius in pixels, that is 59. The whole conversion is done as follows:

*Radius of Target circle $_{image}$ = 59 pixels*

*Radius of Target circle $_{original}$ = 2.2 cm*
*Radius of Target circle $_{image}$ = Radius of Target Circle$_{original}$*
*59 pixels = 2.2cm*
*1 pixel = (2.2 / 59) cm = 0.037 cm*
*Euclidian Distance $_{in\ cm}$ = Euclidian Distance $_{in\ pixels}$ x 0.037*

Next, we measured the distance from the pilot's eye to the stimulus and converted the distance to visual angle. We compared the average error in accuracy in different G values (Figure 10). We did statistical analysis to find any significant difference in accuracy in different G values. We have found a significant effect of conditions $[F(5, 1416) = 102.94, p < 0.05, \eta^2 = 0.27]$. A set of pair-wise t-tests have confirmed significant differences between errors in different G values at $p < 0.05$, though there was no significance difference in errors in different timestamps of 1G.

Next, we measured the distance from the pilot's eye to the stimulus and converted the distance to visual angle. We compared the average error in accuracy in different G values (Figure 10).

We undertook statistical analysis to find any significant difference in accuracy in different G values. We have found a significant effect of conditions $[F(5, 1416) = 102.94, p < 0.05, \eta^2 = 0.27]$. A set of pair-wise t-tests have confirmed significant differences between errors in different G values at $p < 0.05$. There was no significance difference in errors in different timestamps of 1G.

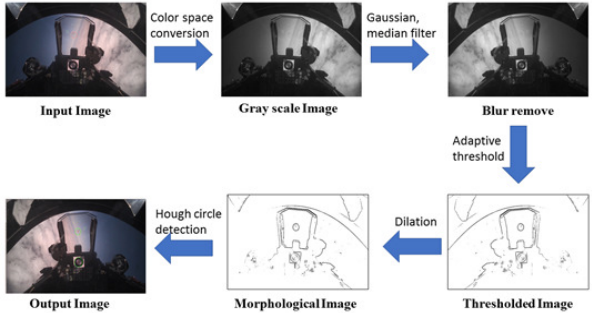

**Figure 8. Different processing steps of the algorithm**

**Algorithm 1** Image Processing Steps

```
function ImagePreprocessing(I)
-------------------------------------------
Input: Input image of size M×N
Output: Preprocessed Image of size M×N
1 Convert the Input Image to gray scale.
2 Apply the Gaussian filter on Image obtained in Step-1
3 Apply the Adaptive thresholding on Image obtained in Step-2
4 Apply the Morphological transformation on Image obtained in Step-3
end
```

**Algorithm 2** Gaze Distance Calculation

```
function AverageGazeDistance (I)
-------------------------------------------
Input: Input image of size M×N
Output: Average Gaze distance between Gaze circle and Target circle
1 Declare empty sets Circles = {} and Gaze_circle_list= {} and Target_circle_list = {} and Flag=0
2. Apply the Hough Transformation on Image and Append all the circle co-ordinates to the Circles List
3. if Circles = None
    then
        Go to End
4 for each (X,Y,R)_{eachcircle} ∈ Circles_{List}:
    if R < Gaze Circle Threshold
        then
            Append the (X, Y, R) to Gaze_circle_list
            Flag=1
    if R < Target Circle Threshold
        then
            Append the (X, Y, R) to Target_circle_list
end
5. if Flag = = 0 and Target_circle_list = = None
    then
        Go to End
6. (x_1, y_1) ∈ Gaze_circle_list and (x_2, y_2) ∈ Target_circle_list
7 Calculate the Euclidean Distance (in pixels)
    Euclidean Distance = $\sqrt{(x_2 - x_1)^2 + (y_2 - y_1)^2}$
8 Calculate the Manhattan Distance (in pixels)
    Manhattan Distance = $|x_2 - x_1| + |y_2 - y_1|$
End
```

We also analyzed different ocular parameters like eye gaze fixations and pupil dilation during different phases of flight. It may be noted from figure 11 that the average fixation duration and fixation rates were both highest during inverted flight at -1G and either fixation duration or fixation rate was higher for 3G and 5G conditions compared to 1G condition.

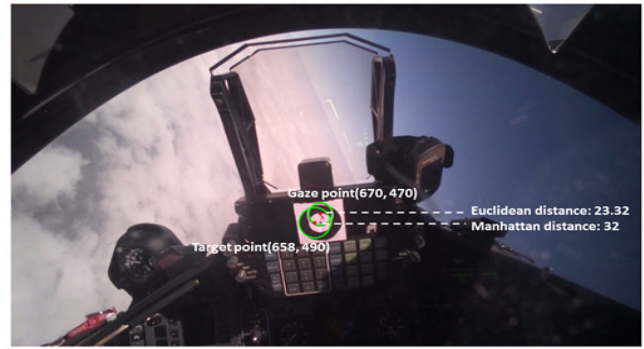

**Figure 9. Evaluation procedure**

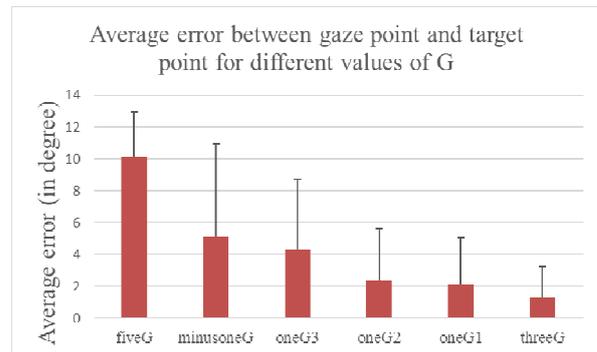

**Figure 10. Average error in gaze accuracy**

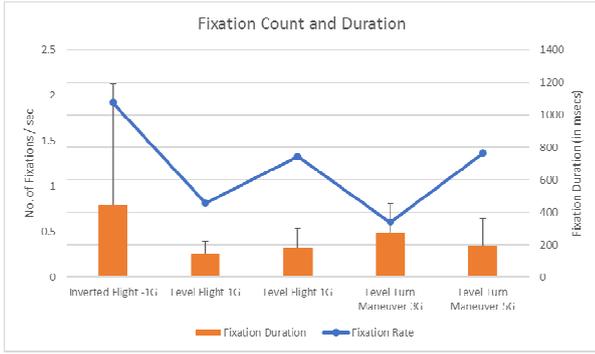

**Figure 11. Fixations at different phases of flight**

In line with our error measurements of COTS eye gaze trackers under various flight conditions, in the next section, we present our analysis on various failure modes of COTS eye gaze trackers in actual fighter aircraft flight conditions.

## 4. STUDY ON FIGHTER AIRCRAFT – ANALYSIS ON FAILURE MODES

We have recorded data from two flights using the COTS eye tracker (Tobii Pro Glasses 2) which uses infra-red (IR) illumination-based eye gaze estimation principles [21]. The duration of the first flight is 55 minutes 58 seconds (Flight 1) and another flight's duration is 56 minutes (Flight 2), the flight profiles are furnished in Table 1 below. The eye tracker contains a front-facing scene camera which records the first person view of the pilot. It also contains four eye-cameras, two cameras per each eye, to record the eye movements. The eye tracker estimates gaze points at a frequency of 100 Hz. The frame rate of scene camera is 25.01 frames/second at 1920 x 1080 resolution and that of each eye camera is around 50 frames/second with a resolution of 240x240. Each gaze point is recorded with a dedicated identifier, called "gidx". We initially used Tobii Pro Lab tool to analyze the recorded gaze samples and observed that both flight recordings contain gaze samples only for around 50% of the duration. We investigated this loss of data samples during the flight using the raw data provided by manufacturer in json format and by correlating the raw data with the eye images.

Table 1 Flight Profiles

| S. No | Objective | Profile |
|---|---|---|
| Flight #1 | Maneuvering flight with head mounted eye tracker on Pilot in Command | Take-off – climb – level flight to Local Flying Area – Constant G (3G and 5G) level turns both sides each – Vertical loop – Barrel Roll – Air to Ground dive attack training missions – Descent – ILS Approach and landing |
| Flight #2 | Non - Maneuvering flight with head mounted eye tracker on Pilot in Command | Take-off – climb – level flight to Local Flying Area – Straight and Level cruise with gentle level turns – Descent – ILS Approach and landing |

The raw data obtained in json format contains various other information recorded during the flight like gyroscope and accelerometer data. We discarded the irrelevant information and retained the data points required for our investigation of lost gaze points.

At first, we synchronized the raw data stream and the eye camera stream in time scale since eye camera stream starts off with an offset from raw data. This is achieved using the Position Time Stamps (PTS) provided in both data streams. We also find that the different frequencies of these two streams is a challenge for data synchronization. Hence, we considered the time duration between two successive frames of eye camera stream and consider all the corresponding gaze data points recorded during that time window. Thus, the latter frame and these data points together form one pair of synchronized data points. Each time windowed raw data may contain multiple gaze points. Every gaze point with its "gidx" contains a status code, 's' which indicates the error associated to that datapoint, if any. The status code 0 indicates no error and any non-zero value of s indicates an error associated is with the data point. We observed that all the gaze points with a non-zero status code are recorded as zeros for both x and y directions [0.0, 0.0]. The gaze points are provided in normalized values; hence the minimum gaze point is [0.0, 0.0] and the maximum is [1.0,1.0].

We segmented the synchronized data points into two categories. The first category *category1* contains eye stream frames whose corresponding gaze points have zero status code. The second category *category2* contains those eye frames with all corresponding gaze points with non-zero status codes. There are frames whose data points have only a subset of gaze points contains zero status code. We did not consider these frames in our analysis as it brings uncertainty on eye image tagging.

For Flight 1, we observed that out of 167,647 frames, only 57,111 frames fall under *category1* and 69,732 frames fall under *category2*. For Flight 2, we observed that out of 167,567 frames, only 81,911 frames fall under *category1* and 51,402 frames fall under *category2*.

Summarizing, 41.6% of the frames does not have any gaze points recorded during Flight 1 and for Flight 2, this stands at 30.7%. Further, if we just look at unsynchronized raw data, both flights recorded more than 51% of the gaze samples are error-prone.

We visually inspected these flight recordings and we hypothesize two reasons for this data loss.
1. Higher levels of illumination on eyes may affect the eye tracker resulting in no gaze estimation.

2. Limited field of view (FoV), especially in the vertical direction, renders the eye tracker with no gaze estimates when user looks beyond the tracking range.

We validated our hypothesis 1 using the eye images in the above mentioned two categories. Since the recorded video stream is an IR video, we converted all eye images into grayscale and computed average of all the pixel values for each image present in both categories. Figure 12 represents the histogram of image intensities for *category1* and *category2* for Flight 1. Figure 13 represents the same for Flight 2. Figure 12a indicates that 93% of the images under *category1* have an average intensity less than 131. But, *category2* contains 42% with average

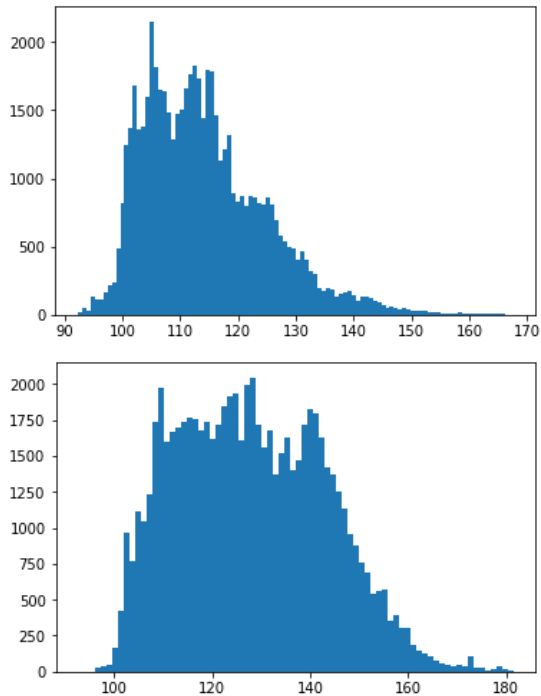

**Figure 12. Histogram of image intensities for Flight 1**

**12a. *Category1*(Top) 12b. *Category2* (Bottom)**

intensity higher than 131. Further, this can also be observed in Flight 2 case, shown in Figure 2. Around 42% in *category2* have higher intensity than 150, while *category1* contains 94% of the images with intensity less than 150. This indicates that images with higher illumination, precisely above 131 in Flight 1 and above 150 in flight 2 have low probability to obtain accurate gaze estimates.

While this evidence supports our hypotheses 1 partially, we observed that there is overlap in the left and right histograms plotted in Fig 12 and Fig 13. Hence, we could not identify a clear average image intensity threshold in order to identify all the failure modes of eye gaze estimation.

We further investigated the data points in *category2* to understand the 58% of the datapoints which have lower image intensities than above mentioned thresholds for each flight using our hypotheses 2. Since we observed that the gaze estimates are lost for a sequence of eye image frames, we clustered the datapoints in *category2* based on their "gidx"s. If a sequence of datapoints under *category2* are having successive gidx's, then all those points are considered as a single cluster. Hence, each cluster can contain one datapoint or several datapoints. Extending our hypotheses 2, we assumed that the pilot must be looking at a

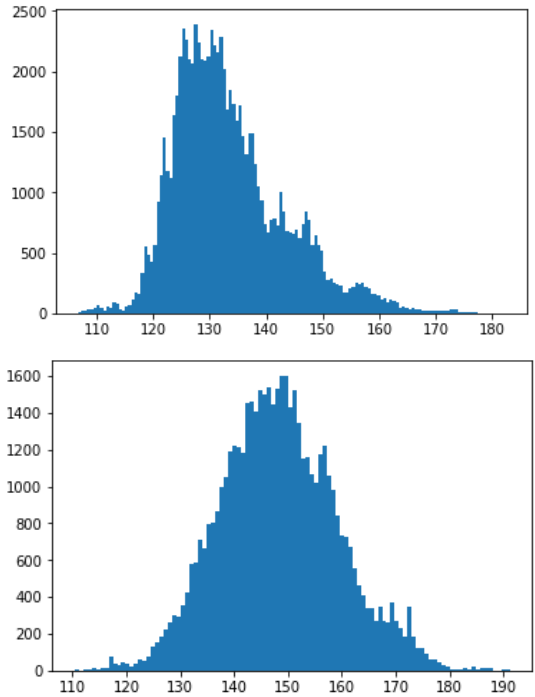

**Figure 13. Histogram of image intensities for Flight 2**

**13a. *Category1*(Top) 13b. *Category2* (Bottom)**

position closer to the extreme tracking positions (beyond which eye tracker cannot track), just before or after the eye tracker fails to provide gaze estimates. During our visual inspection of first person video recorded using eye tracker, we observed that the pilot looks down for various activities like looking at the information displayed in the Multi-functional displays (MFD)s while keeping the head faced horizontal (perpendicular to the vertical axis of the aircraft). During such scenarios, we observed that gaze points were not recorded.

Hence, we analyzed three preceding or subsequent datapoints adjacent to each cluster, which we refer to as boundary datapoints. We looked for boundary datapoints with gaze values beyond 0.8 and less than 0.2 (in both x and y). If any of the boundary datapoints satisfy above criterion, then we may infer that the loss of gaze points is due to the pilot looking beyond the tracking range of the eye tracker.

For flight 1, we obtained 12178 clusters for 69,732

datapoints. For these clusters, 11,865 (97.43%) clusters have boundary points that satisfy the above criterion. To understand image intensities for these datapoints, we plotted a histogram of the image intensities for the datapoints whose boundary points satisfy above criterion. We observed that these image intensities lie in the range of (96,145). This is clearly in the overlap range identified between Figure 12a and Figure 12b.

Similarly, for flight 2, we obtained 8646 clusters for 51,402 datapoints. For these clusters, 8408 (97.24%) clusters have boundary points that satisfy the above criterion. Interestingly here as well, we observed that the histogram of image intensities for the above points lie in the range of (117,164), which is the range of overlap identified in Figure 13a and Figure 13b.

Thus, we infer that, this eye gaze tracker could not identify beyond certain illumination level or if the user is looking beyond its tracking range. We should note that the pilot is performing his assigned tasks during the flight and maintained his natural behavior. This indicates that the tracking range offered by this eye tracker is not sufficient for military aviation environments.

Hence, using our two hypotheses and the raw data, we studied the failure modes of eye gaze tracker in aviation environment. We further add that, while commercial off-the shelf eye trackers may be used in real aviation environments, researchers and practitioners should keep in mind about both the horizontal and vertical tracking range of the eye tracker and it's robustness to external illumination as there is a high chance that the illumination varies rapidly at high altitudes in high speed maneuvers.

## 5. HMDS IN MILITARY AVIATION – EXISTING SYSTEMS

With this understanding of how pilots gaze patterns and the behavior of eye gaze trackers vary over various critical flying tasks, we consider adaptive displays which facilitate eye gaze-based interactions as a solution to reduce the cognitive load of the pilot and the error in interaction. A recent survey [16] with the pilots also suggests that the potential advantages they foresee by using gaze-controlled interfaces are the faster access to information, increased system overview and increased situational awareness. Smyth [17] proposed a system with Heads-Up display (HUD) with electromagnetic head movement tracking and interaction using eye tracker control. However, in order to interact with the HUD, i.e., a display in the line of sight of the pilot, using eye tracking, a pilot has to calibrate initially. If one chooses to interact with HUD using a wearable eye tracker, the gaze points from the wearable eye tracker are obtained with respect to a fixed head position. This affects the performed calibration once the pilot changes his head position or orientation. This issue of interacting with HUD has not been considered in earlier works.

The present BAES Striker II and Elbiit Dash helmet systems provide Head mounted display systems (HMDS) using opto-inertial sensors to track head movement and adapt the content on the display accordingly. These present HMDS also enable pilots to lock on target by head movement. In the case of HMDS interaction, the present systems require more input besides head movement when there is more than one target in a single line of sight. Presently, targets are automatically prioritized, and pilots select the target to engage by using a flip switch. Besides, there can be only so much information that the HMDS display can accommodate without cluttering the visual field. Placing the information on an extended virtual display canvas and facilitating interaction with this content on HMDS can provide a lot more information in a structured and clutter-free manner to the pilot for gaining system overview and situational awareness. In this direction, we propose to use eye gaze directly to engage target while there are multiple targets in one line of sight or/and for interacting with the content on HMDS screen. We proposed an algorithm to integrate both head orientation and eye gaze information into single data which can be used for selecting multiple targets in a given line of sight. With the help of the proposed system, one of the disadvantages perceived in [16], "too much information" can be overcome by facilitating virtually larger canvas and allowing the pilot to customize the information they need on the specific locations of HMDS. The next section describes design and evaluation of such a head and eye gaze movement-controlled system.

## 6. EYE GAZE AND HEAD MOVEMENT-CONTROLLED HMDS – THEORY AND DESIGN

Shree [12] reported the use of an eye-gaze controlled interface in a flight simulator, although that system did not allow user to have free head movement for interaction. In this section, we propose a multimodal eye gaze and head interface, which supports natural head movement along with eye gaze to interact with the HMDS.

*Gaze Direction Vectors*

We used a commercial off-the-shelf wearable eye gaze tracker (Tobii Pro glasses 2) to capture gaze direction unit vectors for each eye. These vectors were measured with center of the respective pupil as the origin. In subsequent sections, we termed left and right eye's gaze direction vectors as *eyeL* and *eyeR* with dimensions 3x1. Wearable eye trackers provide gaze information with respect to a given head position. Figure 14 illustrates this with two instances of user gazing at two different points wearing eye gaze tracker. In the first instance, user looks straight at point A. In the second, he turns his head towards right by α degrees and looks straight at point B. The eye gaze vectors from the eye tracker would be same in these two cases even though the user is looking at two different points in space.

*Head Orientation*

Tobii pro glasses 2 has in-built MEMS (micro-electromechanical system) accelerometer and gyroscope. These MEMS sensors are prone to noise and absence of an in-built magnetometer does not guarantee an accurate measurement of head's yaw [13]. Hence, we used an 9-axis IMU to measure head's yaw, pitch and roll and it is placed right above the user's head. We considered the initial position of user's head as the reference head position. The three mutually perpendicular axes passing through the center of the IMU at this position was considered as the reference coordinate axes. Figure 15 illustrates yaw, pitch and roll with respect to user's head and these are the orientations about axes z, y, x respectively. In subsequent sections, we termed the yaw, pitch and roll of head as α, β, ϒ respectively.

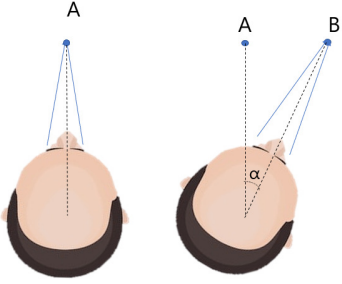

**Figure 12. Illustration of gaze direction vectors along with head movement**

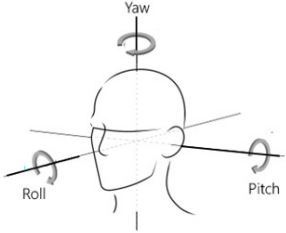

**Figure 13. Yaw, Pitch and Roll for a head movement**

*Head Compensated Gaze Vectors*

Every sample of *eyeL* and *eyeR* obtained along with a given head orientation was transformed to the reference coordinate axes using the following intrinsic 3D transformation [11]. We termed the head compensated gaze vectors as $eyeL_{hc}$ and $eyeR_{hc}$, with dimensions 3x1. Here, we assumed that both eyes are positioned equidistant from the origin of reference co-ordinate axes. The above framework provided unique $eyeL_{hc}$ and $eyeR_{hc}$ for a given point in space. These head compensated eye vectors, $eyeL_{hc}$ and $eyeR_{hc}$ were cascaded into single vector $eye_{hc}$, with dimensions 6x1.

$$T_z(\alpha) = \begin{pmatrix} \cos\alpha & -\sin\alpha & 0 \\ \sin\alpha & \cos\alpha & 0 \\ 0 & 0 & 1 \end{pmatrix} \quad (1)$$

$$T_y(\beta) = \begin{pmatrix} \cos\beta & 0 & \sin\beta \\ 0 & 1 & 0 \\ -\sin\beta & 0 & \cos\beta \end{pmatrix} \quad (2)$$

$$T_x(\gamma) = \begin{pmatrix} 1 & 0 & 0 \\ 0 & \cos\gamma & -\sin\gamma \\ 0 & \sin\gamma & \cos\gamma \end{pmatrix} \quad (3)$$

$$T = T_z(\alpha)T_y(\beta)T_x(\gamma) \quad (4)$$

$$eyeL_{hc} = TeyeL$$
$$eyeR_{hc} = TeyeR$$

We analyzed gaze direction unit vectors for various screen co-ordinates and inferred that the relationship between components of these vectors and screen positions was not linear consistent with previous work [12]. Hence, instead of a linear formulation, we used a 9-point calibration routine to obtain $eye_{hc}$ at different locations on screen and used a feed forward neural network to learn the mapping function.

*Calibration and Operation*

We used an 80" projected display of resolution 800x600 and users were requested to sit at 3.2 m from it. The setup can be viewed at https://cambum.net/Aerospace20.mp4. Users were asked to wear eye tracking glasses and IMU-attached cap (Figure 16). We displayed nine squares of size 90x90 px as calibration markers on screen one after another. Pfeuffer [15] reported limitations of using static calibration markers. To overcome such limitations, we provided visual feedback in response to user's focus on the square. The size of the square reduced continuously when the user focused on it until it reached a minimum size of 10x10 px. We measured standard deviation of gaze vector components to measure the user's focus on the square. In case the user looked away, standard deviation increased and if it were greater than the design threshold, the square regained the original size. This method allows user to stop and resume the calibration at his/her will.

When the squares reached the minimum size, $eye_{hc}$ vectors are recorded for all 9 points. We chose these 9-points in such a way that they span across the screen. We used Tensorflow.NET and Keras.NET for building and training a neural network. A 2 hidden layer neural network was trained with Mean Squared Error loss function and Adam optimizer. We used training loss and coefficient of determination to determine the termination condition of training the network. In addition to that, these parameters were useful in preventing overfitting. The training started once the $eye_{hr}$ vectors were obtained for all 9-points and the average time for neural network training is observed to be 6 seconds. Once the network was trained, predictions of the neural network was used to move the screen cursor.

We performed time window-average of 0.2 sec on input gaze vectors and output predictions to achieve a smooth cursor movement. In addition to that, we also used the following measures to keep cursor less jittery.

- **Pixel Threshold**: We measured the Euclidian distance between successive predictions from the neural network. We updated the cursor position only if that distance were above certain pixel threshold.
- **Angle Step Threshold**: As it is natural to have small head movements, there will be a continuous change in orientation values. In addition to this, IMU has a rms error of 2° and takes 0.2-0.5 seconds to converge to accurate value. Hence, we updated the head orientation only if the incoming orientation value differed from the current value by an angle-step-threshold.

These two design parameters affect both task completion time and jitter in cursor movement. If we set these thresholds too high, small cursor movements cannot be made, and if we set these too low, user might be subjected to irritation with micro cursor movements. Even though the afore-mentioned framework provides direct mapping of $eye_{hr}$ to screen coordinates, error from the IMU affects predictions from the neural network, which results in cursor offset. Participants are able to correct this error by moving their heads while keeping their gaze at the same point on screen.

## 7. EYE GAZE AND HEAD MOVEMENT-CONTROLLED HMDS - USER STUDY

We conducted a user study to compare two interaction modalities, the existing Joystick based TDS and proposed multimodal head and eye gaze interface (MMHE). In the following section we described the flight simulator and design of the study involving various sensors.

*Flight Simulator Setup*

We designed a flight simulator to conduct the user study in dual task setting. Using our setup, participants undertook standard flying tasks in parallel with representative pointing and selection task. This setup allowed us to measure not only pointing and selection times, but also the total response time, consisting of the time required to switch from primary flying to secondary mission control tasks. We designed our study to emulate a Head Mounted Display System (HMDS) where information is projected on to the visual field. Participants needed to interact with that information along with the primary flying task. The flight simulator was projected on to an 80" display.

Third-party flight simulator YSFlight with data logging feature was configured for this study as an F/18 E/F Super Hornet aircraft. The flight simulator was configured with a Thrustmaster Warthog HOTAS (Hands On Throttle and Stick). Both flight simulator and secondary pointing tasks were run on an Intel Pentium CPU G3220@3GHz computer running Windows 7 operating system with 4GB RAM and Nvidia GeForce 210 graphics card.

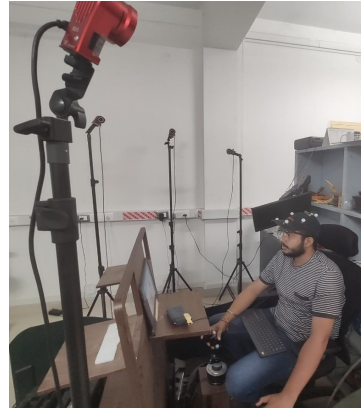

**Figure 14. Participant wearing eye tracking glasses and a cap with IR reflective markers and IMU**

*Head Tracking using OptiTrack*

During the development, we observed that IMU values drifted from actual values which may result in erroneous condition. To study this in detail, we used a COTS infrared based motion capture system (OptiTrack system) to obtain head orientation. We placed 5 retro-reflective markers onto the same cap where IMU was placed. We used 5 Flex 13 cameras to obtain head orientation. We did not use head orientation values obtained from OptiTrack as part of our proposed interface; Rather, we investigated the correlation between the head orientation values obtained from IMU with head orientation from OptiTrack. This setup acts as test-bed and allows us to compare any other head orientation measuring technique apart from IMU in future, that enable us to choose the most accurate head tracking system to integrate with our gaze interface. Since the sampling rate of IMU and OptiTrack is different, we performed time sampling of 1 second to compute the average value. These average values were used to compute correlation.

*Flying Task*

A map was configured with a straight line drawn in the middle. Participants were instructed to fly between 1000 and 2500 feet along the straight line. The secondary task was initiated after the flight reached the designated flight envelope of 1000 and 2500 feet.

*Secondary Task*

We designed Pointing and selection task similar to ISO 9241-9 (Figure 17). This task was overlapped onto the flight simulator and participants undertook this task while flying.

The task was to click a button at the center of the screen followed by clicking a random red color target button. The time between these two clicks was measured as the selection time.

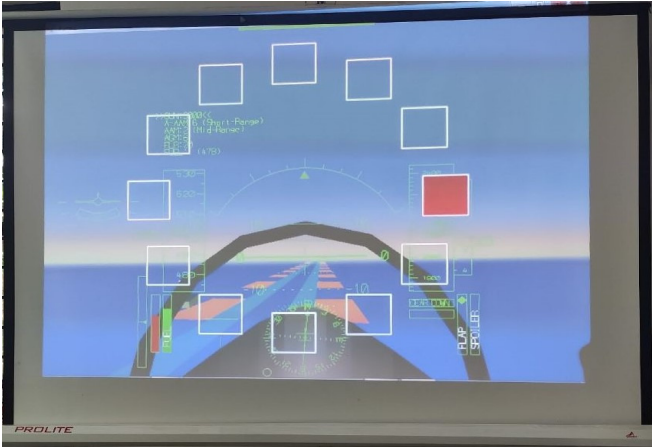

**Figure 15. ISO pointing task overlapped onto the primary flying task**

We considered 3 widths (W) 70px, 80px, 90px, (correspond to 12cm, 14cm and 16 cm) for target buttons and 3 distances (D) between the center button and the target buttons 200px, 220px, 240px, leads to a total of 9 Index of Difficulty (ID) cases. The order of these ID cases and interaction modalities was randomized for all participants. The above-mentioned target widths subtended a visual angle of 2.14°, 2.5° and 2.86° respectively.

In case of Joystick, participant used the trackball on throttle for both pointing and selection whereas in the case of MMHE, participants pointed using their eye gaze or/and head and selected using the button on throttle. The time taken by each participant for the study was recorded since the inception of take-off. The task was considered complete when they performed all 18 clicks using a given modality or when they completed 6 minutes since take-off, whichever was earlier.

We collected data from 8 participants (7 male, 1 female) aged between 23 and 28 years (Mean = 25, SD=1.51). Each participant was instructed to consider flying as the primary task and perform the secondary task only when he/she feels their flight was satisfying the flying task instructions. Each participant performed the task 2 times for each ID case and hence a total of 18 clicks in each mode of interaction. Mean Time (MT) was measured as the average of selection times across all participants for a given ID. ID and Throughput (TP) were calculated based on the following formulae

$$ID = \log_2\left(\frac{D}{W}+1\right) \quad TP = \frac{ID}{MT}$$

All participants were allowed to familiarize themselves with the interface and the actual trial was conducted only after they felt confident in using the system. In the case of MMHE, participants were briefed about the head movements they could perform to look as well as to correct the offset. After each participant completed his/her trial, subjective feedback was collected using NASA TLX for cognitive load and SUS questionnaire for subjective preference. A demonstration of the system can be viewed at https://cambum.net/Aerospace20.mp4

*Results*

Table 2 summarizes both quantitative and qualitative metrics of interaction. We measured mean values of metrics for all participants followed by the standard deviation in parentheses.

**Table 2. Summary of Interaction Metrics**

| Metric | Joystick | MMHE |
|---|---|---|
| Mean Time (MT) (ms) | 4456 (731) | 3017 (909) |
| Throughput (TP) (bits/sec) | 0.434 (0.04) | 0.686 (0.20) |
| TLX Score | 45.63 (15.8) | 37.92 (15.49) |
| SUS Score | 64.68 (14.9) | 73.44 (13.37) |

We analyzed mean times for all Indices of Difficulty for both Joystick and MMHE modalities. The average MT for Joystick and MMHE modalities is 4.5 and 3.0 seconds respectively. Figure 18 shows the MT for all ID cases and the dashed line indicate the trend line for respective modality. We undertook a paired t-test (*t: 4.31, p=0.001, Cohen's d: 1.44*) for MT and found that participants took significantly less time in using MMHE than Joystick.

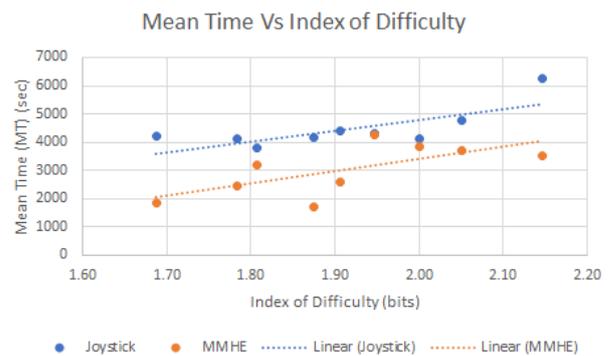

**Figure 16. Mean Time Vs ID for Joystick and MMHE**

The average TLX scores for Joystick and MMHE are 45.63 and 37.92 respectively. A paired t-test (*t: 2.31, p=0.027, Cohen's d: 0.82*) for TLX score indicate that the perceived cognitive load in MMHE case is significantly lower than in Joystick case. Even though the average SUS score for MMHE is higher than joystick, a paired t-test for SUS scores indicate that the subjective preference between Joystick and MMHE is not significantly different.

*Discussion*

In addition to the task completion metrics and qualitative feedback, we analyzed Cursor movement efficiency metrics for both modalities (Figure 19). We used a * mark on the graph for parameters, which were significantly different at p<0.05 in a paired t-test. The cursor efficiency metrics computation took the entire trajectory of the mouse from the center button click till the target button click. The cursor was moving along with eye movement in this dual task setting while using MMHE unlike Joystick case.

We observed participants assessing their flight control by observing the altimeter and relative position with respect to the central path after clicking the center button and before clicking the target button. A significant higher ODC (t: 4.38, p=0.002, Cohen's d: 1.55), MDC (t: 4.73, p=0.001, Cohen's d: 1.68) and higher TAC in MMHE than in joystick can be explained by this observation. The average RE in MMHE and Joystick is 0.85 and 0.34 respectively. MMHE has lower value in metrics that look at the variability of the movement (MV, MO) and it is significantly lower (t: -2.01, p=0.04, Cohen's d: -0.71) than Joystick in terms of movement error (ME). We analyzed the data with Radial Stacked Bar Chart (Figure 20) to obtain better insight about the relation between the above-mentioned metrics and various ID cases. The left and right side of the Radial Stacked Bar chart represents MMHE and Joystick respectively. We can see from the figure that the extent of green, violet and orange representing avg_MO, and avg_ME and avg_MV for all the ID cases on left side is smaller than right side of the chart. We compared participants' flying performance while performing the task with both interaction modalities. Table 3 summarizes the three metrics that represent flying performance. Participants had to fly longer to complete the task when joystick was used than MMHE. The deviation from central path was also higher while using joystick than MMHE. The altitude deviation was not significantly different between two interaction cases.

*Table 3. Summary of Flying Performance*

| Metric | Joystick | MMHE |
|---|---|---|
| Deviation from path | 486.8 (591) | 375.2 (357.1) |
| Altitude Deviation | 199.4 (46.4) | 199.4 (39.7) |
| Average Flight Distance | 56564 | 53313 |

*Table 4. Correlation of Head Orientation between OptiTrack and IMU*

| Participant | Yaw | Pitch | Roll |
|---|---|---|---|
| 1 | 0.83 | 0.82 | 0.23 |
| 2 | 0.92 | 0.72 | 0.76 |
| 3 | 0.93 | 0.67 | 0.38 |
| 4 | 0.85 | 0.78 | 0.44 |
| 5 | 0.75 | 0.88 | 0.43 |
| Average | 0.85 | 0.77 | 0.45 |

We measured head orientation using both OptiTrack and IMU sensors. Out of 8 participants, for 3 participants head orientation data measured from OptiTrack had high error per marker and missing data, hence we could not consider that data. We measured correlation between the values from IMU and OptiTrack for the remaining five participants. We observed high correlation between IMU and OptiTrack for both yaw (0.85) and pitch (0.77) measurements. We observed positive, but low correlation for roll measurements (0.45). While using MMHE, we observed that participants performed offset correction using their head movements when targets appeared in upper quarter of the task region. This might be attributed to a lower correlation of pitch when compared to yaw where pitch movement was required to look at the target.

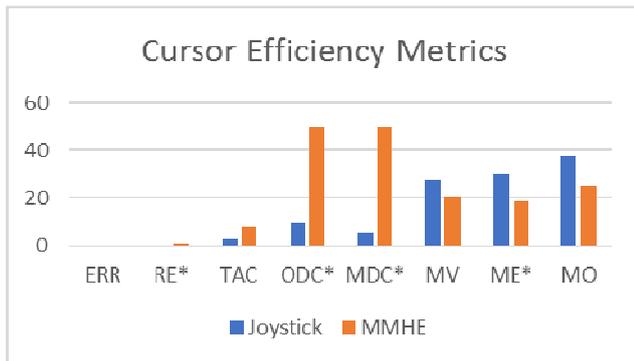

**Figure 17. Comparing cursor efficiency metrics**

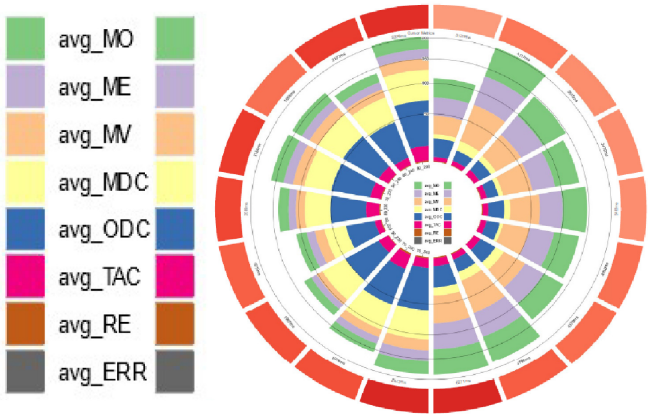

**Figure 20. Radial Stacked Bar Chart**

Thus, we develop and evaluate our multimodal eye gaze and head movement controlled interface on a simulated HMDS. In the next section, we describe screen mounted eye tracker development which can be utilized to operate MFDs in military aviation.

## 8. SCREEN MOUNTED EYE TRACKER FOR MFDS

In this section, we have described three different eye tracking systems and compared them through user studies. Initially we described the different algorithms used for estimating gaze followed by two user studies.

*HoG based Gaze Tracking System*

We used a pre-trained facial landmark detector with iBUG 300-W dataset [22], which works on classic Histogram of Oriented Gradients (HoG) feature combined with a linear classifier to detect facial landmarks [23]. In comparison, Haar cascades are a fast way to detect an object but often detect more false positives compared to HoG and linear classifier [24]. HoG features are capable of capturing the face or object outline/shape better than Haar features. On the other hand, simple Haar-like features can detect regions brighter or darker than their immediate surrounding region better. In short, HoG features can describe shape better than Haar features and Haar features can describe shading better than HoG features. In this case the shape is more important as we need the landmarks of the face hence HoG features produced a better result. After detecting eye region, we detected pupil location by selecting the smallest rectangle possible in the eye region where the pupil can exist. After retrieving pupil locations, we calculated the Eye Aspect Ratio (EAR). We have noted that the eye aspect ratio changes with respect to the distance between the user and the camera. We have modified the EAR calculation formula by using the distance between the two eyes as denominator.

*Webgazer.js*

We implemented a second system using webgazer.js [25, 26] to compare performance of the proposed system. Webgazer.js runs entirely in the client browser. It requires a bounding box that includes the pixels from the webcam video feed that corresponds to the detected eyes of the user. This system uses three external libraries (clmtracker, js_objectdetect and tracking.js) to detect face and eyes. It has methods for controlling the operation which allows us to start and stop it. We have taken the mean of last thirty points from webgazer.js for better target prediction and accuracy of system. We also calculated the mean value during this time to predict the gaze location on a webpage.

*Intelligent System*

We have developed a gaze block estimator which maps user's eye movements to 9 screen blocks using OpenFace [27] toolkit. Since the OpenFace (Figure 21) was reported to have an error of 6☐ for gaze point estimation, we designed a calibration routine which uses the gaze vector data from OpenFace and maps user's eye movements to screen blocks, instead of screen pixels. We have divided the screen into 9 blocks of equal area. We designed a smooth pursuit based calibration routine where a marker traverse across all these 9 blocks and user was asked to follow the marker's movement. The corresponding gaze vectors from OpenFace were recorded and stored with the respective block number as the label. Once the marker completes its path, a neural network is trained to map these gaze vectors to 9 blocks of the screen. For this classification task, we used a 2 hidden layer network with 256 and 128 neurons respectively with cross-entropy loss function and with Adam optimizer. We used the 70% of the data we recorded during calibration for training, 15% for validation and the rest for testing. On a i7 processor computer, we observed that each epoch takes around 0.8 seconds and we trained the network till the test accuracy reaches 90%.

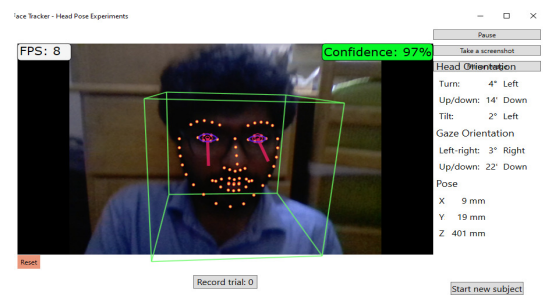

**Figure 21. Screenshot from OpenFace Face Tracker**

*User Study*

We undertook the following user study to compare different eye tracker implementations in different lighting conditions and compared them with a COTS screen mounted eye gaze tracker.

*Participants*

We collected data from 9 participants (8 male, 1 female). All participants were recruited from our university. They do not have any visual or motor impairment.

*Material*

The user trial was conducted on a Microsoft surface pro tablet powered by dual-core processor and it comes with 8 GB RAM and running Microsoft Windows 10 operating system. The surface has a 5 MP camera, which was used to estimate gaze direction.

*Design*

We wanted to use the eye tracker to operate a graphical user interface with limited number of screen elements, hence instead of traditional precision and accuracy measurement, we calculated the pointing and selection times for a set of fixed positions in screen.

We created a user application in which we divided the screen into nine blocks and one of the blocks gets randomly highlighted with blue color as shown in Figure 22a. If the user clicks on the blue block, it turns green as shown in Figure 22b and a different block was highlighted. If the user is unable to click on the highlighted block within 10 seconds, it turns randomly some other block to blue. Using this interface, we calculated the response time by measuring the time difference between appearance of a highlighted block and its selection. Users selected target using the left mouse button.

The trial was performed twice - once in laboratory with lux meter reading 180-200 lux and the other in outdoor condition with lux meter reading between 1800-3000 lux.

The trial consisted of four eye tracking implementations

- HoG based bespoke screen mounted gaze tracker
- Webgazer.js based screen mounted gaze tracker
- OpenFace based intelligent screen mounted gaze tracker
- Tobii PCeye mini eye-gaze tracker (referred hereafter as COTS tracker)

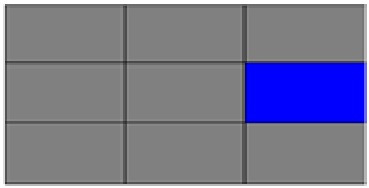

a) Randomly highlighted block in blue

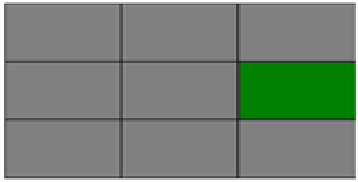

b) After click, highlighted block turns green

**Figure 22. Pointing Task application**

For all trial conditions, we conducted trial on the same user application discussed before. The order of conditions was randomized to minimize practice or learning effect. Each participant undertook all trial conditions.

*Results*

We recorded 322 pointing tasks inside room and 270 pointing tasks outside. We measured the time difference between onset of a target and its correct selection. We removed outliers by identifying points greater than outer fence. We removed only one point from the selection times recorded from the bespoke eye gaze tracker while 12 data points were found values higher than outer fence for the COTS eye gaze tracker.

Figure 23 presents the average selection times and standard deviation. Participants took lowest time to select target using the COTS tracker. We undertook a 2 × 2 unbalance factorial regression based ANOVA (type of eye gaze trackers × lighting conditions) on the response times.

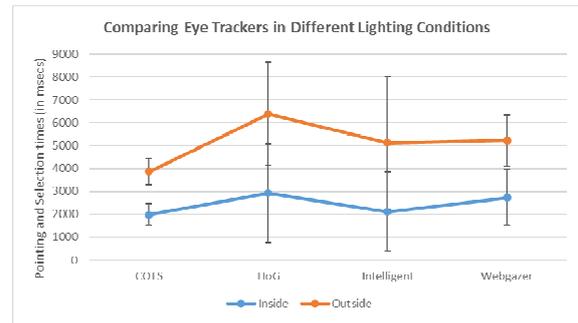

**Figure 23. Comparing Average Response Time among different eye trackers**

We found

- significant main effect of type of eye gaze tracker $F(3,567)=15.44$, $p<0.01$
- significant main effect of lighting condition $F(1,567)=4.05$, $p<0.05$
- significant interaction effect of type of eye gaze tracker and lighting condition $F(3,574)=3.45$, $p<0.05$

Then we undertook two one-way ANOVAs for each lighting condition and found significant main effect of eye gaze tracker implementations.

- Inside room $F(3,318)=8.43$, $p<0.01$
- Outside room $F(3,266)=11.31$, $p<0.01$

Finally, a pair of unequal variance t-tests did not find any significant difference between COTS tracker and our intelligent eye gaze tracker implementation inside room at $p<0.05$ although the difference in response times between the intelligent system and COTS tracker was significant at outside condition at $p<0.05$.

## 9. DISCUSSION

Even though numerous studies were conducted earlier evaluating the utility of eye gaze trackers in military aviation, we conducted multiple user studies in actual flying conditions. These user studies helped us to understand the effectiveness of eye gaze controlled interface in transport aircraft in terms of pointing and selection tasks. Further, our we performed another study to investigate the accuracy of COTS eye trackers under various G-scenarios which occur normally as a part of military flight operations. Our analysis shows that existing COTS sensor can track eye gaze within 4º of visual angle up to +3G and accuracy reduces to 9.5º of visual angle at +5G. This measurement can be used to design future display for eye gaze-controlled HMDS (Head Mounted Display System). However, it may be noted that this paper reports result from only one pilot who has not used eye gaze-controlled interfaces before, and future studies will collect data from more pilots.

Using the data from our studies further, we studied and identified the probable conditions where COTS eye trackers may fall short of expectations in terms of military aviation requirements. We identified that the COTS eye tracker we consider have lesser vertical tracking field than what is necessary in common military aviation operations. We also found that high natural illumination externally may cause failure of IR based eye tracking systems fail to provide gaze estimations. In our current analysis, this illumination levels vary between the two flights under consideration. We may need to analyze more flights data to identify a definitive illumination level beyond the COTS eye tracker may fail. In this direction, we attempted to develop eye trackers in natural illumination conditions using a web camera. We utilized existing state-of-the-art deep learning based eye gaze estimation method to develop an eye gaze controlled interface. Our evaluation of this system along with other screen mounted approaches indicate that these systems can be utilized to perform eye gaze estimation under natural sun light, since they do not rely on IR illumination. The reported task times in Figure 23 indicate that COTS eye tracker perform faster gaze estimation, but this may be due to the fact that COTS eye tracker is built on dedicated optimized hardware, whereas other systems are evaluated on consumer level laptops. In future, we focus to develop gaze point level estimation using deep learning techniques, instead of gaze block estimation from web camera images.

## 10. CONCLUSIONS

This paper reports two user studies on analyzing accuracy of eye gaze controlled interface using COTS eye gaze tracker in military aviation environment. The main findings of our study are
1. Pilots can undertake representative pointing and selection tasks at an average duration of less than 2 secs and less than 5% error rate using eye gaze controlled interface.
2. Accuracy of commercial eye gaze tracking glasses reduces when the constant load factor of aircraft is more than +3G or less than 0G.

However, our studies involved only a limited set of pilots who are not exposed to eye gaze-controlled interface earlier.

We furthermore described a new multimodal head and eye gaze movement-controlled HMDS and compared performance of the system with a traditional Joystick-based TDS in a flight simulator. From our user study, we observed that participants took significantly less time to interact with the targets and perceived significantly less cognitive load using proposed interface than with the existing system. We observed that the cursor movement variation metrics are lower in MMHE than in existing joystick system. In addition to that, participants deviated less and completed the task in shorter distance using our proposed system. These results motivate us to design a multimodal head and gaze interactive head mounted display systems (HMDS). We plan to setup a test-bed for head tracking systems which we plan to develop in the future to integrate with our proposed framework.

We also described the evaluation of our screen mounted eye gaze tracking system against a COTS IR based eye gaze tracker and other web cam based eye gaze tracking approaches. We show that interaction with our system let users to complete the pointing and selection tasks faster than any other screen mounted approaches and we also demonstrate the robustness of screen mounted approaches to external illumination conditions. Our future work includes to develop person-independent eye gaze point estimator systems using deep learning techniques.

### ACKNOWLEDGEMENTS

Authors would like to thank all the participants who took part in the user studies.

**BIOGRAPHY**

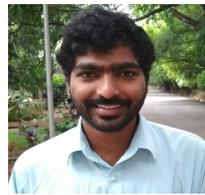

***L. R. D. Murthy***, *received a B.Tech degree in electrical and electronics engineering from SASTRA University, India in 2015. He is currently pursuing a PhD at the Intelligent Inclusive Interaction Design (I3D) lab in the Centre for Product Design and Manufacturing, Indian Institute of Science. His research interests include multimodal human computer interaction, Virtual Reality, machine learning, and artificial intelligence.*

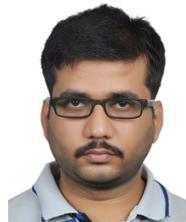

***Abhishek Mukhopadhyay*** *is a PhD student in Computer Science and Engineering at Indian Institute of Information Technology, Kalyani, India. He is working as Junior Research Fellow at Indian Institute of Science. His research aims to explore the possibilities of computer vision and image processing in modern electronic devices accessible to different group of users in different areas.*

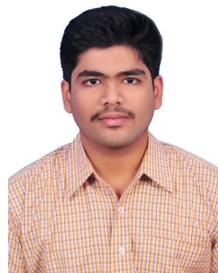

***Yelleti Varshith*** *is currently pursuing his under graduation in Computer Science and Engineering from Indian Institute of Information Technology (IIIT), Kalyani. He is a research Intern at Intelligent Inclusive Interaction Design (I3D) Lab, Center for Product Design and Manufacturing (CPDM), Indian Institute of Science (IISc), Bangalore. His research interests include Machine learning, Deep learning and Artificial Intelligence*

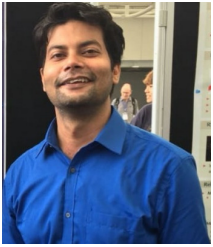
***Somnath Arjun*** *is currently pursuing his PhD at Intelligent Inclusive Interactive Design (I3D) Lab in Center for Product Design and Manufacturing (CPDM), Indian Institute of Science. His research interest includes High Dimensional Data Visualization, Interactive Visualization and Machine Learning.*

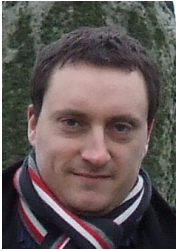
***Peter Thomas*** *is a senior lecturer in aerospace engineering in the School of Engineering and Computer Science, and a member of the Centre for Engineering Research, at the University of Hertfordshire. He delivers courses on flight mechanics, stability and control systems, aeroelasticity, and aerodynamics and his main research interests are in nonlinear flight dynamics and control, and biomechanical systems. After completing his MEng (Warw) in mechanical engineering and a PhD (Cran) in aerospace engineering Peter worked at the Department of Aerospace Engineering in the University of Bristol, working on the ASTRAEA programme for autonomous air-to-air refuelling of unmanned aerial vehicles. He then worked at the Engineering Design Centre in the University of Cambridge, developing evolutionary human-machine interfaces (HMI) for aerospace applications. Peter has also worked on a number of other industrial collaborations including experimental flight testing, bio-inspired flight control, and the implementation of technology-focused methods for engineering education. His current research focus is on developing biomimetic and smart-material-based morphing aircraft.*

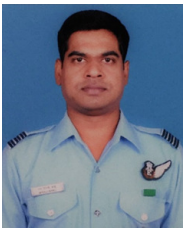
***Wing Commander M Dilli Babu*** *is serving as Flight Test Engineer in Indian Air Force and has flight test experience in over 13 types of fixed wing aircraft including Advanced Combat Aircraft. He has vast instructional experience in flight testing and has been associated with several flight test programs including development of Beyond Visual Range Air – Air Missiles. Presently he has been associated with Experimental Flight Testing of India's First Indigenous Light Transport Aircraft. His research interests include pilot workload evaluation and applications of eye gaze tracking in Aviation.*

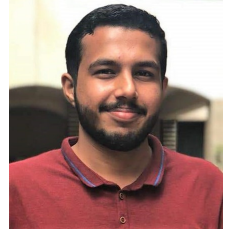
***Kamal Preet Singh Saluja*** *is a researcher at Intelligent Inclusive Interaction Design (I3D) lab, Centre for Product Design and Manufacturing (CPDM), Indian Institute of Science (IISc), Bangalore. He undertook a first degree in Computer Science and Engineering from Rajiv Gandhi Proudyogiki Vishwavidyalaya, Bhopal. He interned at Innovation, Design Study and Sustainability (IDeaS) lab, CPDM, IISc. He is investigating design constraints and developing eye gaze-controlled interaction systems for children with severe speech and motor impairment. He is also working on assessing Cognitive Impairment for elderly people.*

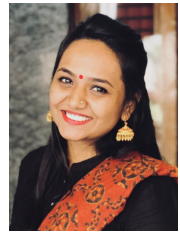
***Jeevithashree DV*** *is a doctoral student from Intelligent Inclusive Interaction Design (I3D) Lab, Centre for Product Design and Manufacturing (CPDM), Indian Institute of Science (IISc), Bangalore. She holds a master's degree in Computer Science & Engineering from VIT University, Vellore, Bachelor's in engineering in Information Science & Engineering from PESIT Bangalore. She has worked as a Junior Research Fellow (JRF) at I3D Lab. She is currently working on investigating Human Computer Interaction issues in limited mobility environment for developing intelligent eye gaze-controlled interfaces. Her research mainly focuses on two distinct user groups- children with Severe Speech and Motor Impairment (SSMI) and military fast jet pilots.*

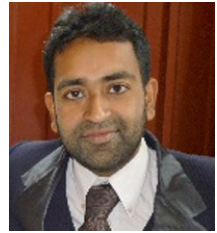
***Pradipta Biswas*** *is an Assistant Professor at the Centre for Product Design and Manufacturing and affiliated faculty at the Robert Bosch Centre for Cyber Physical Systems of Indian Institute of Science. His research focuses on user modelling and multimodal human-machine interaction for aviation and automotive environments and for assistive technology. He set up and lead the Interaction Design (I3D) Lab at CPDM, IISc. He is a Co-Chair of the IRG AVA and Focus Group on Smart TV at International Telecommunication Union. He is a member of the UKRI International Development Peer Review College, Society of Flight Test Engineers and was a professional member of the British Computer Society, Associate Fellow at the UK Higher Education Academy and Royal Society of Medicine.*